# Rolling phase modulation regime for dynamic full field OCT

**Tual Monfort,[1,*] Kate Grieve,[1] Olivier Thouvenin[2]**

*1 Sorbonne Université, INSERM, CNRS, Institut de la Vision, 17 rue Moreau, F-75012 Paris, France*
*[2] Institut Langevin, ESPCI Paris, Université PSL, CNRS, 75005 Paris, France*
*\*tual.monfort@inserm.fr*

**Abstract:** Dynamic full-field optical coherence tomography (DFFOCT) has recently emerged as an invaluable label-free microscopy technique, owing to its sensitivity to cell activity, as well as speed and sectioning ability. However, the quality of DFFOCT images are often degraded due to phase noise and fringe artifacts. In this work, we present a new implementation named Rolling-Phase (RP) DFFOCT, in which the reference arm is slowly scanned over magnitudes exceeding $2\pi$. We demonstrate mathematically and experimentally that it shows superior image quality while enabling to extract both static and dynamic contrast simultaneously. We showcase RP DFFOCT on monkey retinal explant, and demonstrate its ability to better resolve subcellular structures, including intranuclear activity.

## 1. Introduction

Full field optical coherence tomography (FFOCT)[1,2] is a specific configuration of time-domain OCT[3,4], particularly adapted to high-resolution imaging. It captures the 3D distribution of backscattering structures in complex samples with a resolution on the order of 1 $\mu m$ in all dimensions. More recently, dynamic OCT[5–7] and dynamic FFOCT [8,9] have been introduced, which use the temporal quantification of interferometric signal fluctuations to reveal active structures in dense environments. In particular, dynamic (FF-) OCT can reveal single living cells inside biological structures thanks to phase fluctuations associated with active transport of cell organelles[6–8,10]. Dynamic OCT contrast has been shown to relate to metabolic cell activity[8] and to be cell specific[11,12] since organelle transport is a deeply regulated and controlled mechanism at the core of cell physiology[13]. This dynamic contrast can therefore be used as a hallmark for various diseases impacting local cell activity[14–17]. Multiple implementations of dynamic OCT have been described[7], among which dynamic FFOCT (DFFOCT) shows the highest resolution and versatility to be coupled to other imaging modalities[12,18] although axial information cannot be accessed rapidly[4]. Most dynamic OCT implementations rely on calculating metrics that reveal the local strength of OCT signal fluctuations[6–8]. While in the general case, such fluctuations can originate either from intensity or phase changes, OCT is typically more sensitive to axial movements or local refractive index changes by two orders of magnitude compared to transverse displacements[4]. Most previous implementations of dynamic OCT relied on solely natural phase changes arising from the sample without external additional modulation[5,6,8]. Moreover, in FFOCT, it is common to acquire a static and a dynamic image separately. The static image is usually acquired by external phase modulation of the reference field to separate the interference intensity from the incoherent intensity, while the dynamic image is acquired by recording signal fluctuations, induced by the sample, at a fixed optical path length. Nevertheless, as we will show in this manuscript, this approach increases acquisition time, introduces a bias and local phase noise when phase fluctuation amplitude is small, and makes dynamic OCT sensitive to fringe artifacts, and to strong reflectors. This is mainly due to the non-linearity of the cosinus in the interference term. The same phase fluctuation centered on different initial values will result in different intensity fluctuations. Since the phase map in a complex sample is a random map, it creates local noise even at constant reflectivity and phase fluctuation. Whilst schemes to separate the amplitude and phase of the interferometric signal exist, they introduce noise, in particular due to the incoherent nature of light[19], are slower, and produce lower quality dynamic OCT images in practice. Recently, active phase modulation-assisted DFFOCT suggested that higher quality dynamic signal could be obtained by adding an external small-amplitude sinusoidal phase modulation on the order of tens of Hertz in the reference arm[20]. Although this allows the measurement of static and dynamic FFOCT in a single frame, and the normalization of DFFOCT by the local reflectivity, it does not compensate for initial phase variations and still creates noisy images. Additionally, due to the chromatic nature of the phase shift induced by a mechanical motion of the reference arm, this approach would likely fail if large spectral bandwidth is used. In an even more recent paper, S. Morawiec et al. demonstrate dynamic FF-OCM with a high amplitude rapid saw-tooth active phase modulation to improve the SNR of their images[21]. In contrast, in this letter, we describe an alternative implementation of active phase modulation during DFFOCT by slowly linearly increasing the phase over the full duration of DFFOCT acquisition. We will demonstrate that this strategy is optimal to directly access the distribution of local phase fluctuations, improve signal-to-noise ratio (SNR) of DFFOCT images, remove fringe artifacts, reduce speckle, and reveal new subcellular structures. These improvements are demonstrated on freshly excised retinal tissue explants.

## 2. Method description

The setup used is described in Monfort et al.[12]. In short, an LED at 810 nm with a 25 nm bandwidth (M810L3, Thorlabs, Newport, NJ, USA) illuminates a Linnik interferometer using an identical pair of silicon oil immersion objectives with 30X magnification and 1.05 NA (UPLSAPO30XSIR, Olympus, Japan). The reference mirror position can be accurately controlled by a piezoelectric transducer (PK44M3B8P2, Thorlabs, Newport, NJ, USA) piloted by an acquisition card (NI 9263, National Instruments, TX, USA). In total, using a maximum voltage modulation of 10 V, a phase modulation equivalent to up to $7\pi$ can be explored. The same acquisition card is used for synchronizing the camera (Q-2HFW, Adimec, Netherland) and the reference mirror displacement. 512 images were recorded at 100 Hz with a linear stepping of the reference mirror during the acquisition from an equivalent of 0 to either 2 or 4 π, corresponding to 565 pm or 1.13 nm step between each image, respectively. Data was post-processed using MATLAB, using a newly developed metric (see next section) for the brightness channel, and as previously described for hue and saturation channels, once demodulated[12,22].

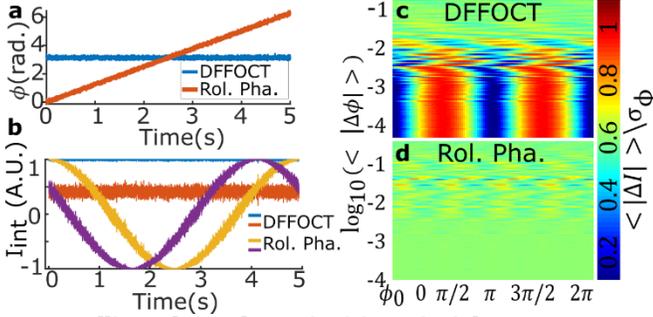

*Fig.1 **Rolling phase dynamic OCT principle.** In contrast to standard dynamic FFOCT where the phase is kept constant during acquisition, in rolling phase DFFOCT (RP-DFFOCT), the reference arm is moved in order to slowly increase the phase (panel a). In presence of sample-induced small phase fluctuations, the interferometric intensity ($I_{int}$) fluctuation strongly depends on the initial phase (0 and $\frac{\pi}{2}$ for blue vs red curves- panel b). In contrast, in RP-DFFOCT, the amplitude of fluctuations becomes independent of the initial phase in rolling phase (yellow vs. purple curves) by homogeneously sampling the full phase space during the acquisition time. Finally, the normalized average intensity variation is plotted for different values of initial phase and different amplitude of random phase fluctuations in log scale in DFFOCT (panel c) and in RP- DFFOCT (panel d) with a reference phase modulation of $4\pi$ in total.*

Adult macaque (*Macaca fascicularis*) retinas were ethically obtained from terminally anesthetized subjects used in unrelated studies, adhering to French Ministry of Education, Higher Education and Research, as well as NIH, and EU guidelines (2010/63/EU). Post-enucleation, 1 cm² retinal samples were embedded in 1% low-melting agarose with Neurobasal-A medium (10888022, Thermo) containing 2 mM of l-glutamine (G3126, MERCK), and sectioned then into 100 μm transverse slices using a vibrating microtome (Leica, Wetzlar, Germany), and prepared for imaging in a glass-bottom plate (Cellvis, P12-1.5H-N, IBL).

## 3. Mathematical formulation

In this section, we will demonstrate that if the linear modulation amplitude is chosen carefully, it becomes possible to measure a metric which depends only on the local reflectivity and the average phase fluctuation amplitude, which are at the core of the dynamic OCT contrast. Let us define $I_r$, $I_s$, $I_{inc}$ as the reference intensity, sample intensity and incoherent intensity respectively. Each scatterer is found at a depth corresponding to a phase $\phi_0$, different for each scatterer, and has a small phase fluctuation $\phi_s(t)$ caused by its intracellular movements. In the rolling phase regime, we add an additional phase variation by slowly moving the reference arm: $\phi_{ref}(t)$. Ideally, $\phi_{ref}(t)$ has a total amplitude larger than $\pi$, and a time scale slower than $\phi_s(t)$. Typically, $\phi_{ref}$ is linearly scanned from 0 to $2\pi$ during the full acquisition (~$N_{frames}$ =512 frames or 5.12 s), $\phi_{ref}(t = k\Delta t) = k\frac{2\pi}{N_{frames}}$.

To simplify, we do not take the coherence length into account, and only consider signal within the coherence volume so that the intensity captured by the camera is written:

$$I(t) = I_r + I_S + I_{inc} + 2\sqrt{I_R I_S}\cos\left(\phi_s(t) + \phi_0 + \phi_{ref}(t)\right) \quad (1)$$

With $Ir, Is, Iinc$ constant. In rolling phase, we calculate the absolute value of the instantaneous intensity difference, which we postulate mostly comes from phase variations (as the scatterers have constant reflectivity, and they do not move outside their initial pixel position at short timescales):

$$|\Delta I|(t) = 2\sqrt{I_R I_S}\left|\cos\left(\begin{array}{c}\phi_s(t+\Delta t) + \phi_0 \\ +\phi_{ref}(t+\Delta t)\end{array}\right) - \cos\left(\begin{array}{c}\phi_s(t) + \phi_0 + \\ \phi_{ref}(t)\end{array}\right)\right| \quad (2)$$

Using trigonometric combinations, we obtain:

$$|\Delta I|(t) = 4\sqrt{I_R I_S}\left|\begin{array}{c}\sin\left(\phi_0 + \overline{\phi_s}(t) + \overline{\phi_r}(t)\right) \\ \cdot \sin(\Delta\phi_s(t) + \Delta\phi_r(t))\end{array}\right| \quad (3)$$

With, $\overline{\phi}(t) = \frac{\phi(t+\Delta t)+\phi(t)}{2}$, and $\Delta\phi = \phi(t+\Delta t) - \phi(t)$

Because $\Delta\phi_r$ is constant and $\sqrt{I_R I_S}\,\Delta\phi_r$ is on the order of the shot noise of FFOCT for most scatterers (since $\Delta\phi_r = \frac{2\pi}{N_{frames}} \sim 10^{-2} \ll 1$), and for $\Delta\phi_s$ typically small, but larger than $\Delta\phi_r$, we can simplify equation (3) to:

$$|\Delta I|(t) \simeq 4\sqrt{I_R I_S}\left|\sin\left(\begin{array}{c}\phi_0 + \overline{\phi_s}(t) \\ +(2k+1)\frac{\pi}{2N_{frames}}\end{array}\right)\right| \cdot |\Delta\phi_s(t)| \quad (4)$$

Hence, this metric is linearly proportional to both the scatterer reflectivity and to the instantaneous phase variation, which is our aim. However, it still depends on the initial phase $\phi_0$. In order to remove this dependency that adds noise to the dynamic images, we want to average this term out using a time average. It then becomes clear that forcing a phase modulation so that the average phase term can explore the full $[0:\pi]$ range is critical so that the sinus term converges towards a constant value. Below, we will discuss different models of phase modulation and scatterer transport to show why and to what extent rolling phase can show better performance than DFFOCT.

**Case 1 : Standard DFFOCT** $\overline{\phi_r} = 0$

From equation (4) without modulation, $|\Delta I|(t)$ strongly depends on $\phi_0$ for samples where the phase fluctuation $\phi_s$ has a low fluctuation amplitude:

$$|\Delta I|(t) \sim 4\sqrt{I_R I_S}\,|\sin(\phi_0)|\cdot|\Delta\phi_s(t)| \quad (5)$$

Figure 1 shows that DFFOCT signal strongly depends on the initial phase $\phi_0$ (x-axis) for phase fluctuations modelled as a normal random distribution of standard deviation up to $2\pi \times 10^{-1.5}$, while becoming independent of $\phi_0$ for larger phase fluctuations sufficient to explore the full phase space.

**Case 2 : Rolling Phase:** $\Delta\phi_r = \frac{2\pi}{N} \ll 1$

From equation (4), although the instantaneous value of $|\Delta I|$ still depends on $\phi_0$, the aim is to calculate the time average of $|\Delta I|$ to average $<\left|sin\left(\phi_0 + \overline{\phi_s}(t) + \overline{\phi_r}(t)\right)\right|>$ to a constant value. It is not directly straightforward that this last term and $\Delta\phi_s$ can be separated, but this appears valid for the most plausible and reasonable hypotheses underlying intracellular transport[13]. We will discuss below 3 regimes of scatterer movements.

In a first regime, $\phi_s$ would be driven by Brownian motion of scatterers (or random noise). $|\Delta\phi_s(\Delta t)|$ can hence be described as a normal distribution of average value 0 and standard deviation $\sigma_\phi$. For such distributions:

$$<|\Delta\phi_s(\Delta t)|> = \sigma_\phi \sqrt{\frac{2}{\pi}}$$

Besides, with $\phi_s$ having an average value of 0, and $\phi\_r$ being scanned between 0 and a multiple of $\pi$, $<\left|sin\left(\phi_0 + \phi_s + \overline{\phi_r}(t)\right)\right|>$ converges towards the average of the absolute value of a sine between 0 and $\pi$, which is equal to $\frac{2}{\pi}$. Then,

$$<|\Delta I|(t)> \sim 4\sqrt{I_R I_S}.\sigma_\phi.\left(\frac{2}{\pi}\right)^{\frac{3}{2}} \sim 0.51\sqrt{I_R I_S}.\sigma_\phi \quad (6)$$

Which is independent of $\phi_0$ for small phase fluctuations, as shown in figure 1d.

In a second regime, $\phi_s$ would be driven by constant directional transport (over the short time course of the experiment), $<|\Delta\phi_s(\Delta t)|>$ is constant, and:

$|\Delta\phi_s| = \left|\frac{2\pi}{\lambda_0} 2v_{zs}\Delta t\right|$. As in the Brownian regime, and by simple simulation, it is possible to confirm that: $1/Texp \int_0^{Texp} |sin(\phi_0 + \overline{\phi_s(t)} + \overline{\phi_r}(t))|dt$ also converges to about $\frac{2}{\pi}$. Because of the directional movement, this time integral corresponds to the integral between 0 and a multiple of $\pi$ plus a small residual. This small residual depends on $\phi_0$ but its standard deviation for different values of $\phi_0$ is always below 4% if the total variation of $\overline{\phi_r}$ exceeds $2\pi$, and is negligible in practice for small scatterers. This residual becomes more and more negligible with increasing strength of $\phi_r$ modulation. However, the instantaneous $\Delta\phi_r$ should be kept small enough otherwise the simplification of equation (3) to (4) is no longer valid, and the static FFOCT image is obtained. In this second regime, the new metric is therefore almost independent of $\phi_0$ in RP DFFOCT and becomes:

$$<|\Delta I|(t)> \sim \frac{8}{\pi}\sqrt{I_R I_S}.|\Delta\phi_s| \sim 0.39\sqrt{I_R I_S}.|\Delta\phi_s| \quad (7)$$

Finally, in the intermediate and more general regime where scatterers have both periods of active transport and periods of Brownian motion, which is described by run-and-scatter models[13], it is not strictly possible to obtain a metric strictly independent of $\phi_0$. However, this has to be mitigated as the persistence time of several common molecular motors exceed the second timescale which is the acquisition duration. Hence, most cases can be described as the first case of Brownian motion with a total phase modulation equal to a multiple of $\pi$ plus a component depending on the sample active transport. As a general rule of thumb, the ideal case would be to roughly match $\phi_r$ exploring at least $2\pi$ during the persistence time of the molecular motor of interest, so that each period of time can be decomposed as a sum of an independent succession of Brownian and directed transport events described above.

## 4. Results

In order to demonstrate the impact of the Rolling phase in combination with the new metric $<|\Delta I|>$, the brightness channel of standard DFFOCT[12,22], calculated as the running standard deviation, is compared to rolling phase (RP) DFFOCT with $<|\Delta I|>$ (Fig2 a vs b).

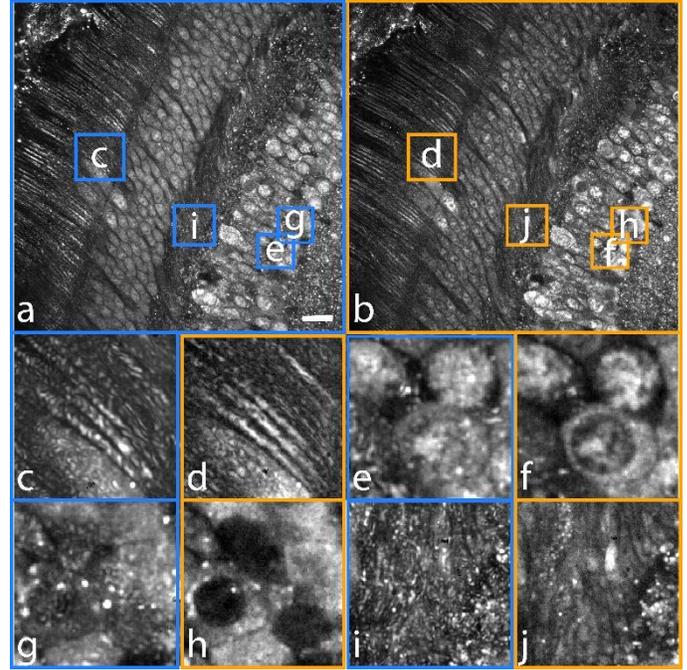

*Fig. 2: Comparison of the brightness channels of DFFOCT (blue) versus Rolling-Phase DFFOCT (orange) on a macaque retinal explant. (a) and (b) show the full DFFOCT (calculated as running STD) and RP DFFOCT (calculated as $<|\Delta I|>$ with $2\pi$ reference phase modulation) images respectively. The retinal region shown in (a-b) runs from left to right through the photoreceptor outer and inner segments, outer nuclear layer, outer plexiform layer, inner nuclear layer and inner plexiform layer. c-j are zoom-ins on a-b, highlighting differences between D-FFOCT (bleu) and Rolling-Phase (orange), respectively. The scale bar is 18 μm for (a-b), 3.5 um for (c-d), 5 μm for (e-f), and 6 μm for (i-j).*

First, fringe artefacts visible around reflective structures are averaged out in RP DFFOCT thanks to the external phase variation (Fig. 2c vs 2d on the photoreceptor outer segments). Second, the global image quality of the brightness channel is improved as phase noise is reduced with RP DFFOCT, enabling new intracellular details to be revealed. In particular, clear nuclear boundaries become detectable with RP DFFOCT (Fig.2e vs. 2f, and 2g vs. 2h), enabling one to distinguish dynamic and heterogenous nuclei (Fig.2d) from uniform nuclei (Fig.2h). Lastly, RP D FFOCT enables a reduction in the global speckle contrast, as observed in Fig.2i-j, here again thanks to the averaging over the full phase space. This reveals axon network boundaries of low contrast (Fig.2i-j). Finally, we note that $<|\Delta I|>$ is much faster to calculate than the running std or other dynamic metrics.[5,23]

A last advantage of RP DFFOCT is that it enables extraction of both static and dynamic signals (Fig. 3) from a single acquisition. A time series of FFOCT is acquired on a retinal explant with standard 2 phase FFOCT (Fig. 3a) for comparison and with RP DFFOCT (Fig. 3c, d). By observing how the intensity is evolving over time on a photoreceptor outer segment, we obtain the profile displayed in Fig.3b, illustrating phase variation over $4\pi$ induced by reference arm movement. Here, two periodic cycles may be observed over the entire acquisition. Extracting the amplitude of the intensity variation using a Fast Fourier Transform (FFT) enables extraction of the static signal (Fig.3c), equivalent to the static "2-phase" image in terms of contrast, but without fringe artefacts (Fig. 3a vs. 3c).

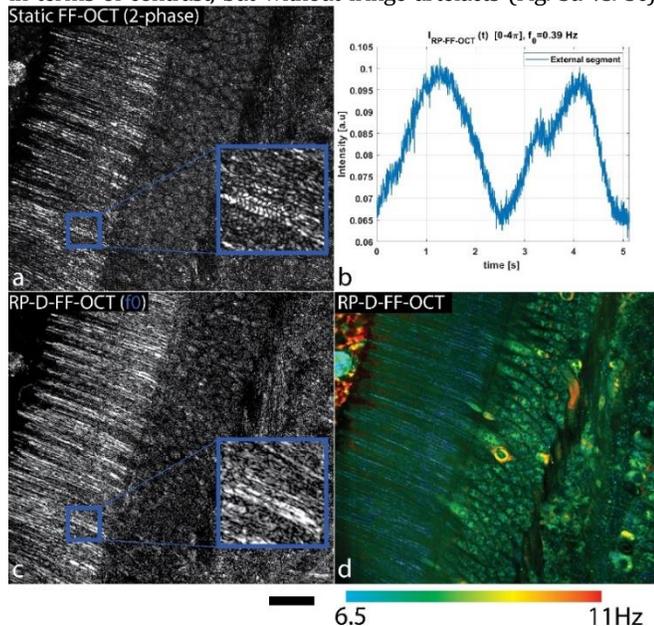

*Fig. 3. RP DFFOCT can extract both static and dynamic components of the interferometric signal in a single shot. (a). Classic 2-phase FFOCT image on a macaque retinal explant. In FFOCT, reflective structures like photoreceptor outer segments show strong fringe artifacts (blue square). (b) In RP DFFOCT on the same sample and region, thanks to the rolling phase of $4\pi$ over the 5.12 s of acquisition duration, the raw interferometric intensity follows a sine wave of carrier frequency ($f_0$ =0.39 Hz). (c) By extracting the magnitude of the Fourier transform at $f_0$, an image of static structures similar to 2-phase FFOCT is obtained, however with reduced fringe artefacts and speckle noise. (d) Dynamic structures can be simultaneously recovered by recombining 3 dynamic metrics. The scale bar is 25 µm and is common to a, c-d.*

Both images in a and c are calculated from the same number of 512 raw images to offer a fair comparison. In order to calculate the Hue and Saturation channels in RP DFFOCT, the mean and the standard deviation of the power spectral density are calculated. However, the dataset must be demodulated prior to calculation, as the beat frequency of the reference phase would mask intracellular signal. To do so, the frequency corresponding to the reference arm movement is simply set to zero, and the power spectral density is recalculated. The brightness channel corresponds to $<|\Delta I|>$ as described above. These three metrics are combined in an HSB image, as shown in Fig.3d to provide direct quantitative observation. Interestingly, with RP DFFOCT, we were able to obtain satisfactory images showing both cellular structure and activity (which typically requires long acquisitions of several seconds), using very little data compared to traditional DFFOCT. With increased SNR, the number of raw images required to distinguish cell types in the retina from the H, S and B maps can be reduced by a factor about 4.

## 5. Discussion

In this work, we have demonstrated that by slowly rolling the phase during the acquisition, we were able to calculate a metric that is almost linearly linked with both scatterer reflectivity and the distribution of phase shifts induced by their transport. This resulted in a dynamic contrast revealing new structures, as well as reducing speckle and fringe artifacts and producing better quality images. Besides, RP DFFOCT enables the capture of both static and dynamic structures in a single acquisition in contrast to standard DFFOCT. Interestingly, rolling phase could be as efficiently applicable to dynamic Fourier domain OCT as well, and should be a quite general result for the OCT community. Finally, the improved SNR reduces the acquisition time of dynamic signals, and calculating $<|\Delta I|>$ as done in RP DFFOCT is much more computationally efficient as usual DFFOCT metrics. This paves the way for faster dynamic measurements, and to visualize dynamic contrast in real time, which is a key parameter for *in vivo* dynamic measurements, or high content screening applications.

**Funding.** The authors wish to acknowledge funding from IHU FOReSIGHT [ANR-18-IAHU-0001], OPTORETINA (European Research Council (ERC) (#101001841)), Région Ile-de-France Sésame, Agence nationale de la recherche (ANR) "OREO" [ANR-19-CE19-0023] and VISCO [ANR-21-CE30-0024].

**Acknowledgment.** The authors want to warmly thank Valerie Forster for the sample preparation, lateral mounting of retinas, and help in writing the biology methods section.

**Disclosures.** "The authors declare no conflicts of interest."